\documentclass[prd, twocolumn, nofootinbib]{revtex4}
\usepackage{epsfig}

\newcommand{\beq}{\begin{equation}}
\newcommand{\eeq}{\end{equation}}
\newcommand{\beqa}{\begin{eqnarray}}
\newcommand{\eeqa}{\end{eqnarray}}
\newcommand{\om}{\Omega_m}
\newcommand{\calr}{{\cal R}}

\def\la{\mathrel{\mathpalette\fun <}}

\def\fun#1#2{\lower3.6pt\vbox{\baselineskip0pt\lineskip.9pt
  \ialign{$\mathsurround=0pt#1\hfil##\hfil$\crcr#2\crcr\sim\crcr}}}

\begin{document} 

\title{Probing Gravitation, Dark Energy, and Acceleration} 
\author{Eric V.~Linder} 
\affiliation{Physics Division, Lawrence Berkeley National 
Laboratory, Berkeley, CA 94720} 

\begin{abstract} 
The acceleration of the expansion of the universe arises from 
unknown physical processes involving either new fields in high 
energy physics or modifications of gravitation theory.  It is 
crucial for our understanding to characterize the properties of 
the dark energy or gravity through cosmological observations and 
compare and distinguish between them.  In fact, close consistencies 
exist between a dark energy equation of state function $w(z)$ and 
changes to the framework of the Friedmann cosmological equations 
as well as direct spacetime geometry quantities involving the 
acceleration, such as ``geometric dark energy'' from the Ricci 
scalar.  We investigate these interrelationships, including 
for the case of superacceleration or phantom energy where the fate 
of the universe may be more gentle than the Big Rip. 
\end{abstract} 

\maketitle 

\section{Introduction} \label{sec.intro}

The acceleration of the expansion of the universe poses a 
fundamental challenge to the standard models of both particle 
physics and cosmology.  In both cases addition of an unknown 
physical component, called dark energy, or modification of 
gravitation, possibly arising from extra dimensions, is required. 
Most attention has been paid to dark energy as a high energy scalar 
field, a physical component contributing a presently dominating 
energy density, characterized by a time varying equation of state. 
But acceleration is fundamentally linked to gravitation through 
the Principle of Equivalence and changes to the framework of the 
Friedmann cosmological equations governing the universal expansion 
would play a natural role. 

Observations from next generation cosmological probes will map the 
expansion history $a(t)$ at 1\% precision, offering the possibility 
of characterizing the physics responsible for the acceleration. 
This can be used to test specific models inspired by unified 
physics involving string theory, supergravity, extra dimensions 
(e.g.\ braneworlds), or scalar-tensor gravity, say.  Alternately, 
one can derive general parametrized constraints on the expansion 
history and propagate these through into quantities such as an 
effective dark energy equation of state, extra terms in the 
Friedmann equations, or spacetime geometry characteristics. 

Not only the magnitude of the constraints but the interpretation 
of them is important.  We investigate to what extent one can use 
a common parametrization to describe these very different areas of 
new physics, and conversely how they can be distinguished.  In 
\S\ref{sec.w} we briefly review dark energy as a scalar field 
component of the universe.  A general modification of the 
Friedmann equation is analyzed in \S\ref{sec.dh}.  We examine in 
\S\ref{sec.rh} the fundamental and general relation between 
acceleration and spacetime geometry, specifically involving the 
Ricci scalar, to motivate modifications of 
gravitation as a possible source of the acceleration -- ``geometric 
dark energy''. 
In \S\ref{sec.super} we address the issue of 
superacceleration and whether this leads to a Big Rip.  We conclude 
in \S\ref{sec.concl}, with thoughts on future prospects for 
understanding how cosmological observations will 
lead us to specific new physics. 

\section{Physical dark energy} \label{sec.w} 

With the discovery of the acceleration of the cosmic expansion 
\cite{perl,riess}, physicists tended to interpret this in terms of 
a new physical component of the universe -- dark energy -- 
possessing a substantially negative pressure.  This is 
perhaps not surprising since models involving the cosmological 
constant had been under consideration and the effects of 
generalized pressure to energy density ratios, or equations of 
state, on cosmological observations had been worked out, e.g.\ 
\cite{wag86,lin88,fpoc}.  Yet, as is well known, the cosmological 
constant can be viewed as belonging to either the right hand, 
energy-momentum tensor, side of Einstein's field equations or to 
the left hand, spacetime geometry or gravitation side.  
Still, in 
analogy to inflation theory, the observations were treated as 
a high energy physics scalar field $\phi$ with a potential 
$V(\phi)$, often called quintessence. 

We here briefly review the essentials so as to later compare and 
contrast the treatment of gravitation as the source of the 
acceleration.  Dark energy as a physical component possesses an 
energy density $\rho_\phi$ and pressure $p_\phi$, both generally 
functions of time $t$, or equivalently cosmic scale factor $a$ or 
redshift $z=a^{-1}-1$. The equation of state ratio is defined to 
be $w_\phi(z)=p_\phi/\rho_\phi$.  The cosmological constant is 
special in possessing $w_\phi=-1$, which ensures that its density 
and pressure are constant in both time and space. 

Like the matter or radiation components of the universe, dark 
energy is generically globally homogeneous and isotropic.  
However, in order for dark energy to dominate the energy density 
of the universe today, but not in the past, in accordance with 
observations, it must have an effective mass $m\sim 
\sqrt{V_{,\phi\phi}}\sim H_0\sim 10^{-33}\,{\rm eV}$, where $H_0$ 
is the expansion rate today, the Hubble constant.  This implies 
that on scales smaller than the horizon size the dark energy is 
smooth and unclustered, while on larger scales it possesses 
inhomogeneities.  This clumpiness is important observationally in only 
restricted circumstances, such as for the growth of matter density 
perturbations on near horizon scales. 

For cosmological observations of the expansion history, e.g.\ 
distances and cosmography, and of the growth of matter 
perturbations on subhorizon scales, the dark energy is simply 
characterized by its energy density $\rho_\phi$ (equivalently 
its fractional contribution to the critical energy density 
$\Omega_\phi(z)=8\pi\rho_\phi/3H^2$) and equation of state ratio 
$w(z)$.  The evolution of the energy density follows 
\beq 
\rho_\phi(a)=\rho_{\phi,0}\,e^{3\int_a^1 d\ln a\,[1+w(a)]}, \label{eq.rhow}
\eeq 
so only the equation of state ratio and the present density enter. 
For a spatially flat universe, the present dimensionless dark 
energy density is related to the matter density by $\Omega_\phi= 
1-\Omega_m$.

From the equation of state function one can recreate the high 
energy physics Lagrangian of the field in terms of its potential 
and kinetic energies: 
\beqa 
V(\phi) &=& \frac{1}{2}(1-w)\,\rho_\phi \label{eq.vw} \\ 
K &=& \frac{1}{2}\dot\phi^2 = \frac{1}{2}(1+w)\rho_\phi 
\label{eq.kw} \\ 
\phi(a) &=& \int da\,\frac{1}{\dot a}\dot\phi=\int d\ln a\, 
H^{-1}\sqrt{2K}, \label{eq.phiw}
\eeqa 
where the last line allows translation from the expansion factor 
to the value of the scalar field.  Thus $w(a)$ really is the 
central, determining quantity. 

Note that the equation of motion of the field $\phi$, the Klein-Gordon 
equation, follows easily from the continuity Friedmann equation: 
$\dot\rho_\phi=-3H(\rho_\phi+p_\phi)=-6HK$.  Since 
$\dot\rho_\phi=\dot K+\dot V=\dot\phi \ddot\phi+V'\dot\phi$, 
where prime denotes a derivative with respect to the field, 
we obtain the relevant equation 
\beq 
\ddot\phi+3H\dot\phi=-V'.
\eeq

It is often convenient to devise a tractable and model independent 
method of assessing the ability of specific models to reproduce 
the observations.  Parametrization of $w(a)$ in a two dimensional 
phase space suits this well; there exist many possibilities but 
one of the simplest, 
\beq 
w(a)=w_0+w_a\,(1-a), \label{eq.wa}
\eeq 
has good success in fitting a variety of scalar field theories, 
especially those with slow variation (of order the Hubble time) 
in the equation of state. 
While there is no requirement that the scalar field partakes of 
the characteristic time scale of the Hubble expansion, many 
classes of models do.  Furthermore, a reasonable fit to $w(a)$ is 
only truly needed over the limited redshift range when the dark 
energy has significant dynamical influence, so eq.\ (\ref{eq.wa}) 
is widely applicable. 

For a best fit, $w_a$ is often taken to correspond to the time 
variation in the equation of state at redshift $z=1$, approximately 
when dark energy is expected to become significant.  That is, 
$w'=-dw/d\ln a|_{z=1}=w_a/2$.  One could also imagine using a 
different ``pivot redshift'' to define $w_0$ and $w_a$, perhaps that 
at which the two parameters are decorrelated.  However in a coarse 
sense this is still mathematically equivalent to eq.\ (\ref{eq.wa}) 
and in a fine sense this disrupts the model independence of the 
parametrization in that the pivot location will depend on the 
specific model and on the cosmological method of probing it. 

The theory of deriving constraints on the dark energy equation of 
state from a variety of cosmological probes has been well addressed, 
including aspects of parameter degeneracies and probe 
complementarity, as well as optimization of observations 
(e.g.\ \cite{fhlt,huttur,hutstark,somehu,linsl}).  Data from next 
generation precision cosmology surveys, for example KAOS 
\cite{kaos}, LSST \cite{lsst}, Planck \cite{planck}, SNAP 
\cite{snap}, etc., should be plentiful and in complementarity 
capable of determining $w_0$ and $w_a$ within $1\sigma$ 
uncertainties of roughly 0.05 and 0.15 respectively. 

Key clues to the fundamental physics responsible for the 
acceleration lie in whether $w_0$ is more negative, more 
positive, or consistent with the value $-1$ and whether $w_a$ 
is negative, positive, or consistent with zero.  Measurements 
consistent with $w_0=-1$, $w_a=0$ would provide circumstantial 
support for a cosmological constant origin, perhaps simply because 
it is the simplest model, but would also give motivation to look 
for large scale inhomogeneities in the scalar field since those, 
possibly in the guise 
of a sound speed $c_s^2<1$, would provide a definitive distinction 
from the cosmological constant.  Of course conversely, values 
incompatible with the cosmological constant do not rule out its 
existence, only that its potential energy must be smaller than 
that of the dominant scalar field. 

Even with tightly constrained values of a few characteristics of 
the equation of state function, such as $w_0$ and $w_a$, we will 
not narrow the field to a specific model.  Most potentials have 
multiple parameters and can cover a swath of such a phase space. 
What the forthcoming observations will tell us is that certain 
classes of models are restricted to some parameter range, and 
other classes are restricted to another parameter range (possibly 
approaching the limit of a simpler model, such as the cosmological 
constant).  
Naturalness and motivation by theory will be needed to winnow the 
results to a theory of new physics. 

But have we been overly narrow in our expectations, by interpreting 
the observations in terms of a physical component arising from 
high energy physics?  Might the acceleration instead signal new 
physics from a change in the form of the cosmological expansion 
equations rather than a change in the ingredients going into them?

\section{Modifications of the Friedmann Equations} \label{sec.dh}

Looking to extensions of general relativity for an explanation 
of the accelerating expansion has several attractive features. 
It does not require introduction of hypothetical scalar fields 
(e.g.\ quintessence), yet may possess close ties to high energy 
physics such as string theory or extra dimensions; it does not 
obviously suffer from fine tuning problems necessarily (e.g.\ 
the Ricci scalar naturally evolves; development of density 
nonlinearities could induce backreaction on the expansion); and 
it is eminently testable by a number of independent cosmological 
measurements. 

\subsection{General Approach} \label{sec.dh.genl}

To test the framework of our cosmology theory we should impose 
prior expectations of the form of a modification as lightly as 
possible.  We have good evidence for the presence of matter 
density in the universe, from both baryons and dark matter, 
neither of which can accelerate the expansion, and 
strong evidence from the cosmic microwave background anisotropy 
measurements that the universe is consistent with being spatially 
flat.  Taking that as the extent of our knowledge, we can 
parametrize our ignorance of the physical cause of acceleration 
with an arbitrary additional term in the Friedmann expansion rate 
equation: 
\beq 
H^2/H_0^2=\om\,(1+z)^3+\delta H^2/H_0^2. \label{eq.hdh}
\eeq 

Note that such a phrasing is more general than a parametrization 
in terms of the matter density exclusively, such as $H^2=f(\rho)$. 
While the latter can easily be reduced to the form of eq.\ 
(\ref{eq.hdh}) by means of taking $\delta H^2=f(\rho)-8\pi\rho/3$, 
the converse is not true.  Indeed, the $f(\rho)$ approach cannot 
deal with simple time varying dark energy models with nonzero 
$w_a$. 

Linder \& Jenkins \cite{linjen} showed that the general form 
eq.\ (\ref{eq.hdh}) was mathematically 
equivalent to a time variable dark energy equation of state function 
\beq 
w_{\rm DE,eff}(z)\equiv -1+\frac{1}{3}\frac{d\ln(\delta H^2/H_0^2)} 
{d\ln(1+z)}, \label{eq.wdh}
\eeq 
as far as cosmography.  That is, observations of the expansion 
rate and distances alone could not distinguish between these possibilities. 
This degeneracy might be broken, however, through other information such 
as the growth rate of matter density perturbations, as discussed below. 

In addition to the effective equation of state we can write down other 
effective ``high energy physics'' characteristics of the modified gravity 
theory.  The total equation of state of the universe follows immediately 
from the continuity Friedmann equation, $\dot\rho=-3H(\rho+p)$, to give 
\beq 
w_{\rm T,eff}(z)\equiv -1+\frac{1}{3}\frac{d\ln(H^2/H_0^2)} 
{d\ln(1+z)}. \label{eq.wtdh}
\eeq 
The corresponding potential and kinetic energies of the effective field 
come from eqs.\ (\ref{eq.rhow})-(\ref{eq.kw}): 
\beqa 
V &=& \frac{3}{8\pi}\delta H^2-\frac{H_0^2}{16\pi}\frac{d(\delta H^2/ 
H_0^2)}{d\ln(1+z)} \\ 
K &=& \frac{H_0^2}{16\pi}\frac{d(\delta H^2/ 
H_0^2)}{d\ln(1+z)}. 
\eeqa 
Note that this is useful as well for treating dark energy models with 
multiple fields; if there are two components (after all, if we discover 
that $w\ne-1$ this does not guarantee there is not still a cosmological 
constant present) then the effective equation of state is a weighted 
average, 
\beq 
w_{\rm DE,eff}=w_1\frac{\delta H_1^2}{\delta H_1^2+\delta H_2^2}+w_2 
\frac{\delta H_2^2}{\delta H_1^2+\delta H_2^2}, 
\eeq 
where $\delta H_i^2$ is the energy density of the $i$th field. 

Various extensions of the Friedmann equation have been considered in 
the literature.  For example, \cite{dvaliturner} consider a term 
$\delta H^2\sim H^\alpha$, motivated by infinite scale extra 
dimensions -- a ``bulk'' encompassing our 4-d ``brane''. 
We initially examine two gravitational source models that lie toward 
the extremes of 
present data on the equation of state.  The first model is the 
extra dimensional braneworld ``leaking gravity'' model \cite{ddg}. 
Here the modification to the Friedmann equation arises from a 
crossover length scale related to the 5-dimensional Planck mass; 
on larger scales the gravitational force felt in our 4-dimensional 
brane is reduced.  This typically has an effective equation of 
state more positive than $-1$ (and corresponds to $\alpha=1$ above). 
The second is the vacuum metamorphosis model of \cite{parkerraval}, 
originating from a convergent sum of quantum vacuum contributions 
of a light scalar field coupled to the Ricci scalar curvature. 
This very elegant approach leads to a rapidly evolving effective 
equation of state that is more negative than $-1$.  To the extent 
currently possible these models have some definite physical 
motivation for their modifications. 

We mentioned above that kinematics, i.e.\ cosmography, would not 
distinguish between these or other such modifications and dark energy, 
by virtue of eq.\ (\ref{eq.wdh}), but that dynamical probes such as the 
growth of structure might break 
this degeneracy.  Let us investigate this further, both in general 
terms and with the specific models mentioned.

\subsection{Role of Complementary Probes} \label{sec.dh.bw} 

For the growth of structure, it is not only the characteristics of the 
global, homogeneous and isotropic, universe that enter but the more 
microphysical properties of the components themselves.  Thus sound speed 
of the dark energy field or interactions with dark matter could give 
information separate from that contained within the equation of motion 
governing the cosmic expansion.  However, for our present case, we are 
trying to distinguish modifications of gravity from canonical physical 
dark energy; if we restrict ourselves to gravitation models obeying the 
Principle of Equivalence and minimally coupled to matter and nothing else 
then there are 
no such microphysical parameters that could break the degeneracy.  
Then all that enter the perturbation growth are the Hubble drag term 
depending on $H(z)$ and the dynamical evolution of the matter density, 
also determined by $H(z)$.  For a contrasting view of the braneworld 
model with a time varying Newton's constant, see \cite{lue}. 
If the dynamics is limited in this way we 
should expect that if we define a modified Friedmann equation and 
associated effective equation of state as in eqs.\ (\ref{eq.hdh}), 
(\ref{eq.wdh}) 
then we cannot distinguish the gravitational origin from the particularly 
crafted dark energy model.  

However, note that while there is a formal correspondence between a 
modification $\delta H^2$ 
and an equation of state $w(z)$, one might expect the resulting function 
to be so complicated that one would be reluctant to ascribe it to a 
physical dark energy.  On the other hand, the modification may be amenable 
to quite a simple dark energy fit.  We examine this for our two test 
models. 

For a flat braneworld model, the crossover scale $r_c$ defines an effective 
energy density $\Omega_{bw}=(1-\om)^2/4 =1/(4H_0^2r_c^2)$ and 
\beq  
\delta H^2/H_0^2 = 2\Omega_{bw}+2\sqrt{\Omega_{bw}}\sqrt{\om(1+z)^3+ 
\Omega_{bw}}.  
\eeq 
The cosmography in the form of the supernova magnitude-redshift 
relation is 
excellently fit\footnote{Here and in the rest of the paper we mean 
specifically that the dark energy model reproduces the modified 
gravity results to within 0.01 mag over the redshift range $z=0-2$.} 
by the simple dark energy model of $(w_0,w_a)=(-0.78,0.32)$.  We take 
both models to have the same matter density, $\om=0.28$. 

For the vacuum metamorphosis model, the cosmic expansion causes 
the quantum vacuum to undergo a phase transition at a redshift 
$z_j$ away from the matter dominated behavior. 
So the modification to the Friedmann equation goes from zero at high 
redshift to 
\beq 
\delta H^2/H_0^2 = (1-m^2/12)(1+z)^4+m^2/12-\om(1+z)^3, 
\eeq 
for $z<z_j$, where $z_j=[m^2/(3\om)]^{1/3}-1$ and 
$m^2=3\om[(4/m^2)-(1/3)]^{-3/4}$. 
For $\om=0.28$, $m^2=10.93$ and $z_j=1.35$.  Despite the rapid 
evolution in the effective equation of state, the magnitude-redshift 
relation is excellently fit by $(w_0,w_a)=(-1,-3)$.  Note that this is 
a physical model for an effective phantom energy, i.e.\ where $w<-1$. 

\begin{figure}[!hbt]
\begin{center} 
\psfig{file=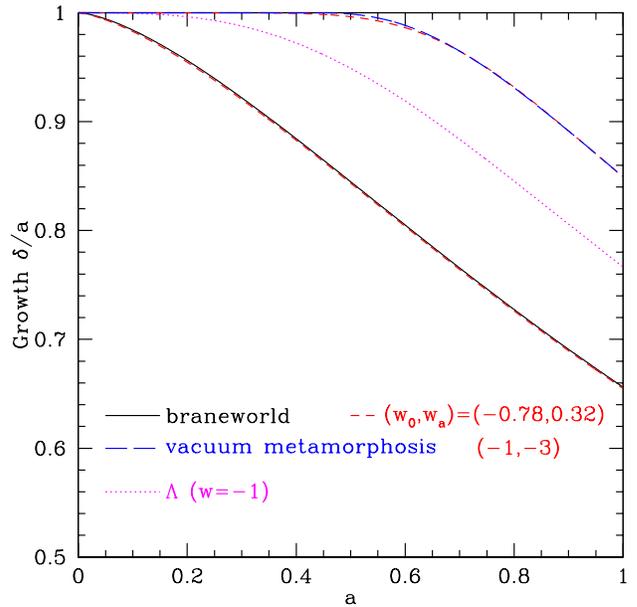,width=3.4in} 
\caption{The growth factor behavior $\delta/a$ for two 
modified gravitation models is compared with that of dark energy models. 
A clear distinction can be seen relative to the cosmological constant, 
$\Lambda$, model, but simple time varying dark energy models (short 
dashed, red curves) can be found that reproduce the modified gravity. 
} 
\label{fig.grobwvm}
\end{center} 
\end{figure}

Does the dynamical probe of the growth of matter density perturbations 
preserve the degeneracy between the gravitational source and the high 
energy physics (dark energy) source for accelerating expansion?  Figure 
\ref{fig.grobwvm} emphatically affirms this.  While the growth evolution 
of either of the models is readily distinguishable from a cosmological 
constant universe, the models cannot be separated from their dark 
energy counterparts.  (One could equally well have first fit 
the growth history and then looked for deviations in the magnitude-redshift 
curves.) 

Note that the braneworld scenario, with its more 
positive equation of state, shuts off growth earlier since its influence 
on the expansion was greater at early times, while the vacuum 
metamorphosis model shows increased growth even compared to the 
cosmological constant case, as generically expected for $w<-1$ models. 
Recall that the linear mass power spectrum is proportional 
to the square of the growth factor, so the models differ $\sim25\%$ 
in power amplitude from the cosmological constant. 

\begin{figure}[!hbt]
\begin{center} 
\psfig{file=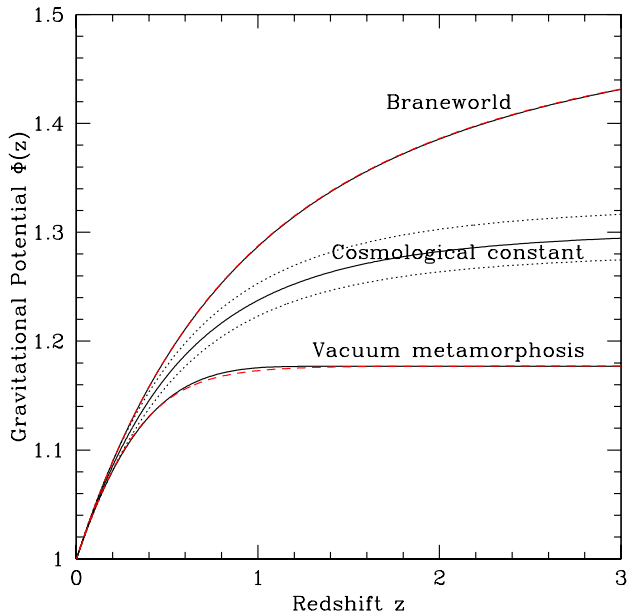,width=3.4in} 
\caption{The gravitational potential $\Phi(z)$ for the same models as 
Fig.\ \ref{fig.grobwvm} is plotted vs.\ redshift, showing the decay 
of the potential as the expansion accelerates.  Dashed, red curves 
are for the mimicking $(w_0,w_a)$ models.  The dotted outliers to the 
cosmological 
constant curve show the deviation expected by a misestimation of the 
matter density $\om$ by 0.02. The discrimination of modified 
gravity from a cosmological constant is clear, but from the fit dark 
energy models is problematic.  
} 
\label{fig.potlbwvm}
\end{center} 
\end{figure}

If we normalize to the present amplitude of structure (this would 
roughly correspond to normalizing the power spectra of the different 
models by the present mass variance $\sigma_8$ rather than to the high 
redshift CMB power) the situation does not change.  Figure 
\ref{fig.potlbwvm} plots this in the form of the gravitational potential 
of the mass perturbations.  Again the gravity and dark energy models 
lie virtually on top of each other.  To indicate a measure of the 
ability of cosmological observations to distinguish models, for the 
cosmological constant case we show the effect of variation in $\om$ 
by $\pm0.02$ (dotted lines).  

The parametrization in terms of dark energy variables $w_0$, $w_a$ is 
nearly the simplest possible, but it is highly successful in mimicking 
the more complicated gravitational modification.  The possibility of 
discriminating between dark energy and gravity would be even worse 
for either a more complicated dark energy ansatz or a nonparametric 
analysis in terms of the expansion history $a(t)$ or density history 
$\rho(t)$ directly.  Correlations between cosmological quantities 
tend to dilute the precision of the nonparametric approach relative to 
the equation of state fit by roughly a factor of 2, 
e.g.\ 0.02 mag or 1\% distance measurements reconstruct the expansion 
history to only 2\% precision \cite{linprl}. 

Another possible cosmological probe is the CMB temperature power 
spectrum.  This is primarily dependent on dark energy or low 
redshift modifications of the Friedmann equation through the 
geometric quantity of the distance to the last scattering surface. 
However it is generally not nearly as sensitive to the equation of 
state as the supernova magnitude-redshift data.  In any case, the 
distances to the last scattering surface agree between each gravity 
model considered and its corresponding dark energy version to 0.1\%, 
below what Planck will be able to achieve. 

\subsection{Discrimination from $\Lambda$} \label{sec.dh.cases} 

While the degeneracies exhibited between the two gravity models and 
their dark energy matches are quite interesting, 
data favors an effective equation of state closer to $w=-1$.  However, 
the braneworld model can only supply this for matter densities 
$\om\ll1$.  For $\om=0.28$ its rough, averaged equation of state is 
$\bar w\approx-0.7$ while that for vacuum metamorphosis is $\bar w 
\approx-1.3$.  Suppose future data continues to narrow in around the 
value $w=-1$; are there gravitational modifications that may be 
confused with a cosmological constant fit? 

We devise additional terms $\delta H^2$ such that they mimic dark 
energy near the cosmological constant value.  These modifications 
to the Friedmann expansion equation are essentially ad hoc, though 
they bear some functional resemblance to physics models such as braneworld 
and $k$-essence tachyon field scenarios (cf.\ \cite{scherrer,chimento}). 
\beqa 
{\rm Case\ 1:}\qquad H^2&=&(8\pi/3)\rho+\sqrt{A'+B'/\rho} \label{eq.case1} \\ 
{\rm Case\ 2:}\qquad H^2&=&(8\pi/3)\rho+\sqrt{A'+B'\rho} \label{eq.case2} \\ 
{\rm Case\ 3:}\qquad H^2&=&(8\pi/3)\rho+\sqrt{A'\rho+B'/\rho}. \label{eq.case3}  
\eeqa 

These are universes with matter density as the only physical component 
dynamically important today, but with modifications to the Friedmann 
expansion equation.  By evaluating these expressions at the present, 
one derives an expression for the constant $A'$ in terms of $B'$ and $\om$, 
so there are only two free parameters.  It is convenient to define 
a dimensionless quantity, $B=B'(8\pi/3H_0^6)$ in Cases 1 and 3 or 
$B=B'(3/8\pi H_0^2)$ in Case 2.  

Case 1 has the property that the effective dark energy equation of state 
ranges between $w\in [-3/2,-1]$.  At high redshifts, $w\to -1$ and today 
$w(0)=-1-B/[2\om(1-\om)^2]\approx -1-3.4\, B$.  For case 2 the range 
is $w\in [-1,-1/2]$, with the value evolving from $w=-1/2$ at $z\gg1$ to 
$-1+0.27\,B$ today.  The relatively large value of $w$ at early times 
is likely to interfere with structure formation.  Very roughly, Cases 1 
and 2 are milder versions of the vacuum metamorphosis and braneworld 
scenarios, respectively. 

Case 3 is intriguing in that $w\in [-3/2,-1/2]$, crossing the cosmological 
constant value of $-1$.  Thus one might imagine that this model could 
mimic on average the cosmological constant at recent times -- and is worth 
studying 
in detail.  Unfortunately, the transition between its asymptotic values 
is quite sharp owing to the difference of six powers of the scale factor 
in the two terms in the square root.  One could adopt a wholly ad hoc 
model containing $\sqrt{A'\rho^\alpha+B'\rho^{-\beta}}$ but we would 
likely learn little physics motivation.  Instead we keep Case 3 as is 
and use it as an interesting, if extreme, test case to investigate model 
degeneracy.  Because of its rapid transition, if this model can be well 
fit by a simple dark energy model then much less radical forms likely 
will be as well. 

In this phenomenology we walk a fine line: if $w>-1$ by too much, the 
model will be uninteresting since it is easily ruled out by observations, 
but if $w\approx-1$ then the modification is too strongly degenerate with 
simple physical dark energy models to probe physics well.  Observations 
are less stringent on ruling out models with $w<-1$, so these are useful 
to explore further, and if they cross through the interesting $w=-1$ 
value then their time averaged equation of state may well satisfy future 
constraints.  Thus Case 3 allows investigation of the extent to which 
distance and growth probes can break degeneracies between classes of 
physics responsible for the acceleration. 

We first consider for which values of the parameter $B$ we can fit the 
data for the least sensitive dark energy probe: the CMB measurement of 
the distance to the last scattering surface.  If we require the distance 
to match the distance in the cosmological constant case to a certain 
precision, then we obtain upper limits to $B$ in Cases 1 and 2, and a 
range in Case 3.  This is because in Cases 1 and 2 the value $B=0$ 
corresponds exactly to the cosmological constant, so these cases will 
never be fully ruled out under our assumption that the true model 
is that of the cosmological constant $\Lambda$.  However Case 3 is distinct 
from a $\Lambda$ model throughout its parameter space. 

\begin{figure}[!hbt]
\begin{center} 
\psfig{file=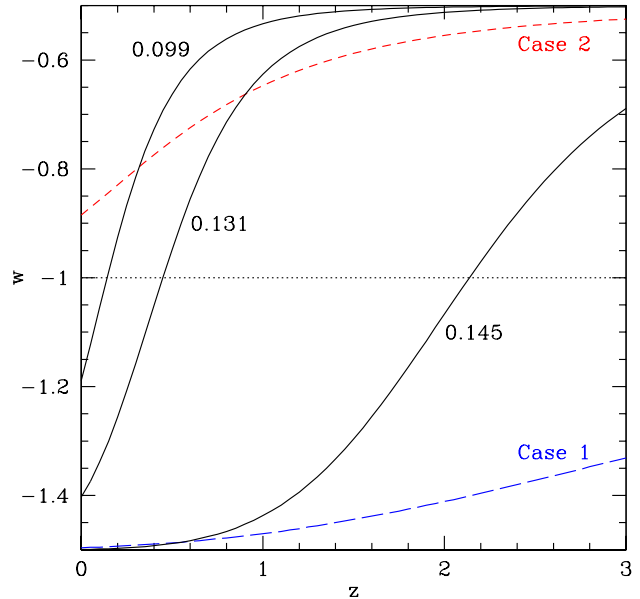,width=3.4in} 
\caption{The effective equations of state corresponding to the 
modified Friedmann equations (\ref{eq.case1}-\ref{eq.case3}) are plotted 
vs.\ redshift.  The parameter space allowed under CMB constraints for Cases 
1 and 2 lie between the respective curves shown and the $w=-1$ line, i.e.\ 
they can mimic a cosmological constant arbitrarily closely.  Case 3 
curves (labeled by value of $B$) can fit the CMB distance of the 
$\Lambda$ model with much more 
strongly varying equations of state, lying between the left and right 
solid curves, with a perfect fit given by the middle solid curve.   
} 
\label{fig.dhw.lss33}
\end{center} 
\end{figure}

Figure \ref{fig.dhw.lss33} illustrates the allowed parameter space for 
the case of WMAP precision: the last scattering surface distance 
$d_{lss}$ known to 3.3\% (1$\sigma$).  For Case 1, the area between the 
long dashed 
curve and the dotted curve at $w=-1$ is allowed, corresponding to 
$B<0.427$.  For Case 2, the area between the short dashed curve and the 
dotted curve at $w=-1$ is allowed, corresponding to $B<0.144$.  Note 
that in both cases the allowed effective equations of state are fairly 
slowing varying functions of redshift, so we expect ease in fitting them 
to a $(w_0,w_a)$ dark energy model and difficulty in discrimination from 
the cosmological constant with whatever probe for $B\ll1$. So we will 
not consider them further.  In Case 3, the 
CMB data would 
restrict the model to have $0.099<B<0.145$, with a perfect match of 
the distance for $B=0.131$.  Nevertheless, the equations of state 
clearly do not resemble that of the cosmological constant, and have a 
strong time dependence. (Note that Planck precision of 0.7\% would 
limit $B$ to between 0.126 and 0.135).  The CMB distance to last scattering, 
normally thought fairly insensitive to time variation, can 
put tension on regions of parameter space for these time varying 
models.

Next we apply the supernova magnitude-redshift and growth tests to the 
models given by Case 3 and see to what extent these can distinguish 
the gravitational model from the cosmological constant, or from 
the best fitting effective dark energy model.  All of these 
gravity models can be distinguished from the cosmological constant, 
$\Lambda$ model through the magnitude-redshift probe; the magnitude 
differences range from 0.1-0.2.  It is not easy to mimic the 
cosmological constant behavior with a modification $\delta H^2$ 
except as $\delta H^2\to\Lambda$.  

However this is a separate issue from whether the modification matches 
{\it some} dark energy model.  In general the degeneracy between a 
gravitational source and effective dark 
energy model remains.  We find excellent 
fits by the simple $(w_0,w_a)$ parametrization as follows: 
$B=0.099$ corresponds to $(w_0,w_a)=(-1.12,1.2)$, $B=0.131$ to 
$(-1.49,1.64)$ , and $B=0.145$ to $(-1.52,0.2)$.  Recall that 
$w'\approx w_a/2$.  The model that exactly reproduces the $d_{lss}$ 
for the $\Lambda$ ($w'=0$) model has a $w'\approx0.8$!  

\begin{figure}[!hbt]
\begin{center} 
\psfig{file=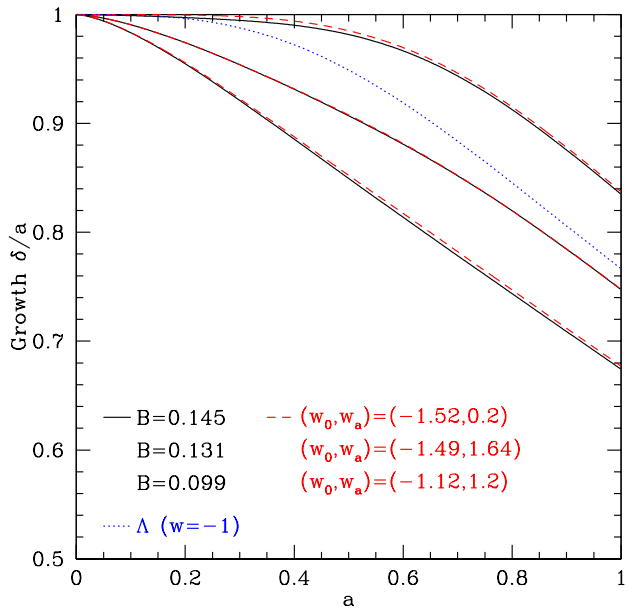,width=3.4in} 
\caption{As Fig.\ \ref{fig.grobwvm} but for Case 3 modified gravity. 
A fairly clear distinction in growth behavior exists relative to the 
cosmological constant model, but not with respect to each corresponding, 
simple, time varying dark energy (dashed, red curves).  These were chosen to 
match the magnitude-redshift relation, so neither expansion history nor 
growth history here distinguishes between a gravitational and dark energy 
explanation for the acceleration of the universe. 
} 
\label{fig.grodhw}
\end{center} 
\end{figure}

Again, the growth of matter perturbations does not break the 
degeneracy, as seen in Figure \ref{fig.grodhw}.  Gravity models 
can be distinguished from each other, and dark energy models from 
each other, but the mapping to the effective equation of state 
holds firm.  Note that the growth behavior of the models from 
Cases 1 and 2 with the largest $B$ values allowed by the CMB data 
roughly agree with the extremes plotted for Case 3.  This implies 
that if CMB data is consistent with the cosmological constant then 
the growth behavior should lie in the region between the upper and 
lower growth curves (at least for the three case forms considered). 

\begin{figure}[!hbt]
\begin{center} 
\psfig{file=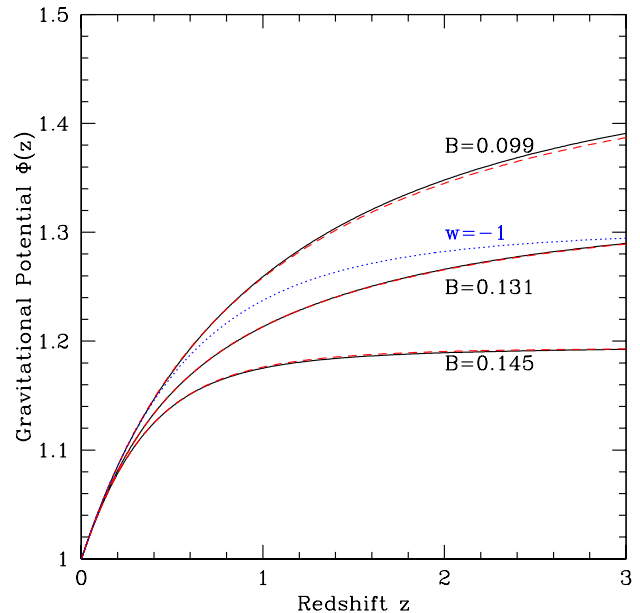,width=3.4in} 
\caption{The gravitational potential behavior as in Fig.\ \ref{fig.potlbwvm}, 
but for the Case 3 modified models (black solid curves) and dark energy 
models (red dashed curves, blue dotted curve for $w=-1$) in 
Fig.\ \ref{fig.grodhw}. 
} 
\label{fig.dhpotl}
\end{center} 
\end{figure}

Differences between the $B=0.131$ (exact match in $d_{lss}$) model 
and the cosmological constant amount to less than 8\% in the power 
spectrum, so the magnitude-redshift data would be the most incisive 
probe.  In Figure \ref{fig.dhpotl} we again normalize to the present 
matter power spectrum and plot the gravitational potential decay 
behavior.  Even such an extreme modified gravity model as the rapidly 
varying Case 3 
cannot be distinguished from a dark energy parametrized by 
$(w_0,w_a)$.  (Note that in fitting $(w_0,w_a)$ models we impose 
$w(z)\le-0.5$ in the growth equation to match the allowed equation 
of state range of the $B$ models, but 
this in fact does not affect the results very much.) 

\section{Acceleration Directly} \label{sec.rh} 

Through the Principle of Equivalence, acceleration has a very 
direct relation to the nature of gravitation and to the spacetime 
geometry.  In turn, mapping the expansion history and observations 
of cosmological distance relations, or cosmography, has a clear 
connection to the spacetime geometry.  This allows future data to 
directly constrain modifications of general relativity, testing the 
framework of the gravitation theory not merely the ingredients of 
the universe.  It seems useful to try to 
make this connection between the measurements and theory as 
explicit as possible, especially in the hope of distinguishing 
a gravitational origin for the acceleration of the expansion of the 
universe from a physical dark energy origin. 

\subsection{Principles} \label{sec.accprin} 

Starting with Robertson-Walker metric for a homogeneous and 
isotropic universe, and imposing spatial flatness, leads to the 
relation between the expansion factor $a(t)$ and the spacetime 
geometry quantity of the Ricci scalar curvature R: 

\beq 
R=6\left(\frac{\ddot a}{a}+H^2\right), \label{eq.rha}
\eeq 
where $H=\dot a/a$ is the Hubble parameter.  No dynamics, i.e.\ specific 
$a(t)$ relation or physical theory, is assumed.   Cosmography directly 
probes $a$ and its derivatives, and hence the quantities $R$ and $H$. 
It is possible that these are not the whole story, that the gravitational 
action contains other terms and so the interpretation of the observations 
in terms of the theory of gravity is more complicated, but so long as the 
metric holds, then the relation (\ref{eq.rha}) is still good.  (See 
\cite{carroll} for the case of $R^{-n}$ terms in the action.) 

Observations of acceleration, $\ddot a>0$, then inform us about the Ricci 
scalar.  In particular, acceleration imposes the condition 
\beq 
R>6H^2. 
\eeq 
Again, this is wholly equivalent at this level to an effective total 
equation of state parameter for the universe, 
\beq 
w_{\rm T,eff}\equiv\frac{1}{3}\left(1-\frac{R}{3H^2}\right). \label{eq.wt}
\eeq 
Indeed we see that $R>6H^2$ corresponds to the usual condition $w<-1/3$. 
The use of $w$ is purely a symbolic definition and does not rely on a 
physical link that would come from, e.g., employing the relation 
$R=8\pi T$ between the Ricci scalar and the trace of the energy-momentum 
tensor that general relativity provides. 

Note that consistency holds between the two approaches of this section 
and \S\ref{sec.dh}.  In some sense we have modified the acceleration 
($\ddot a$) Friedmann equation here and the velocity ($\dot a^2$) 
Friedmann equation in the previous section.  To demonstrate consistency, 
start with eq.\ (\ref{eq.wdh}) and substitute in eq.\ (\ref{eq.hdh}). 
Using the identity 
\beq 
(H^2)\dot{}=4H^3[(R/(12H^2)-1], \label{eq.doth}
\eeq 
one obtains 
\beq 
w_{\rm DE,eff}(z)=\frac{1}{3}\frac{H^2}{\delta H^2}\left(1-\frac{R}{3H^2} 
\right). 
\eeq 
Finally, since the total equation 
of state of the universe is related to the effective dark energy, or 
``parametrized ignorance'', equation of state by $w_T(z)=w_{\rm DE,eff}(z) 
\Omega_{\rm DE,eff}(z)$, we find 
\beq 
w_T=w_{DE}(\delta H^2/H^2)=\frac{1}{3}\left(1-\frac{R}{3H^2}\right), 
\eeq 
as in eq.\ (\ref{eq.wt}).  One can only go from eq.\ (\ref{eq.wt}) to 
eq.\ (\ref{eq.wdh}), however, if one defines an appropriate split between 
knowledge and ignorance, i.e.\ the $\om$ and $\delta H^2$ terms. 

The generality of the link of the total equation of state 
with the spacetime geometry and 
the dynamical eq.\ (\ref{eq.doth}) has an exciting implication. 
The equations point up the centrality of the variable $\calr\equiv 
[R/(12H^2)](z)$, 
since both the equation of state and the Hubble expansion parameter 
can be defined in terms of it.  That is, through eq.\ (\ref{eq.doth}) 
$H$ is determined by 
\beq 
\frac{H^2}{H_0^2}=e^{4\int_0^{\ln 1+z} d\ln y (1-\calr)}. 
\eeq 
Knowledge of the spacetime quantity ${\cal R}$ therefore allows 
us to solve for $H$, $R$, the comoving distance $r(z)=\int dz/H$ and others, 
the magnitude-redshift relation $m(z)=5\log[(1+z)r]$, etc.  This is 
a powerful simplification.  

Furthermore, we will see in \S\ref{sec.super} 
that $\calr=1$ is a critical value, corresponding to $w_T=-1$ (a 
deSitter state) and a 
universe on the cusp between ordinary acceleration and superacceleration. 

\subsection{Parametrization} \label{sec.accpar}

As with the equation of state $w(z)$, forthcoming observational data will not 
be strong enough to reconstruct directly the entire function, here 
${\cal R}(z)$.  Instead we must learn about the physics encoded in it, 
whether gravitational or high energy, in smaller steps.  Following the 
equation of state we might try to parametrize ${\cal R}$ in different 
models by a fitting form containing a few parameters.  Suppose in 
analogy to eq.\ (\ref{eq.wa}) we write 
\beq 
\calr=r_0+r_1(1-a), \label{eq.r01} 
\eeq 
with $r_0$ representing the present value and $r_1$ giving a measure 
of its time variation.  This seems a reasonable minimal parametrization 
for the same reasons as with $w(a)$; one might expect that the spacetime 
geometry should be slowly varying with the expansion. 

In this ansatz, the Hubble parameter is 
\beq 
H=H_0\,a^{2(r_0+r_1-1)} e^{2r_1(1-a)}. 
\eeq 
If we want to ensure a matter dominated epoch at high redshifts ($a\ll1)$, 
then we require $H$ to asymptotically vary as $a^{-3/2}$, thus 
\beq 
r_0+r_1=1/4. \label{eq.rcon} 
\eeq 
This leaves us with only a one parameter family and so we could elaborate 
the fitting form (\ref{eq.r01}) to allow a second parameter.  However, 
we find that for redshifts $z\la 2$, where most of the cosmological probe 
data will lie, the linear fit is a superb approximation to a wide 
variety of physical dark energy models -- as long as the constraint 
condition eq.\ (\ref{eq.rcon}), unnecessary at these redshifts, is not 
imposed.  However one could certainly be fancier and attempt a fit that 
both satisfies the moderate redshift fitting and the asymptotic constraint, 
such as $\calr=1/4+r_0a\tanh(r_1 a)$ or $\calr=1/4+r_2a^2+r_3a^3$ (i.e.\ 
a cubic polynomial with the zeroth order term fixed by the matter 
domination asymptote and the first order term fixed by the smooth approach 
to this asymptote; thus we are left with two free parameters).  But the 
linear fit suffices, matched smoothly to a matter dominated asymptote 
for high redshift calculations. 

Finally, if we have a specific function $\calr$ then we can derive the 
corresponding dark energy model, or its effective equivalent, upon imposing 
a split between matter and dark energy, i.e.\ choosing $\om$.  The 
effective dark energy equation of state is then 
\beq 
w_{\rm DE,eff}(a)=\frac{1}{3}(1-4\calr)\left[1-\om 
e^{-\int d\ln y\,(1-4\calr)}\right]^{-1}, 
\eeq 
and the scalar field potential and kinetic energies follow from 
eqs.\ (\ref{eq.rhow}-\ref{eq.phiw}) as before.  Explicitly, 
\beqa 
V &=& (1+2\calr)\frac{H^2}{8\pi}-\frac{3H_0^2}{16\pi}\om\,a^{-3} \\ 
K &=& (1-\calr)\frac{H^2}{4\pi}-\frac{3H_0^2}{16\pi}\om\,a^{-3}.
\eeqa

\subsection{Distinction from Dark Energy} \label{sec.accde}

As we carried out previously for the Friedmann modifications 
$\delta H^2$, we can investigate the discrimination between this 
direct acceleration, or ``geometric dark energy'', model and physical 
dark energy for various 
cosmological probes.  Once again the straightforward parametrization 
of dark energy in terms of $(w_0,w_a)$ provides an excellent fit to 
the geometry model (note that this is {\it not} a 
consequence of the similar forms of eqs.\ (\ref{eq.r01}) and 
(\ref{eq.wa}), due to the presence of matter; furthermore, the fit 
is similarly successful when using $\calr=1/4+r_2a^2+r_3a^3$). 

We examine four dark energy - Ricci geometry pairs.  For the 
cosmological constant, the fit is provided by $(r_0,r_1)=(0.81,-0.73)$, 
for the time varying equation of state SUGRA model with 
$(w_0,w_a)=(-0.82,0.58)$ the analog is $(r_0,r_1)=(0.69,-0.58)$, 
and for $w=-0.8$ and $w=-1.2$ they are $(0.71,-0.533)$ and 
$(0.92,-0.98)$ respectively.  Each pair possesses magnitude-redshift 
diagrams agreeing within 0.01 mag out to $z=2$. 

Dynamical aspects within the matter density perturbation growth equation 
still contain no leverage to break the degeneracy in any substantial way. 
For reference we write the growth equation of a linear matter density 
perturbation $\delta=\delta\rho/\rho$: 
\beqa 
G'' &+& (3+2\calr)a^{-1}G' \nonumber \\ 
&+& [1+2\calr-(3/2)\om a^{-3}/(H^2/H_0^2)]a^{-2}G=0, 
\eeqa 
where $G=\delta/a$ is the normalized growth and prime denotes a derivative 
with respect to scale factor $a$.  (The growth equation given a modification 
$\delta H^2$ is written in \cite{linjen}.) 

\begin{figure}[!hbt]
\begin{center} 
\psfig{file=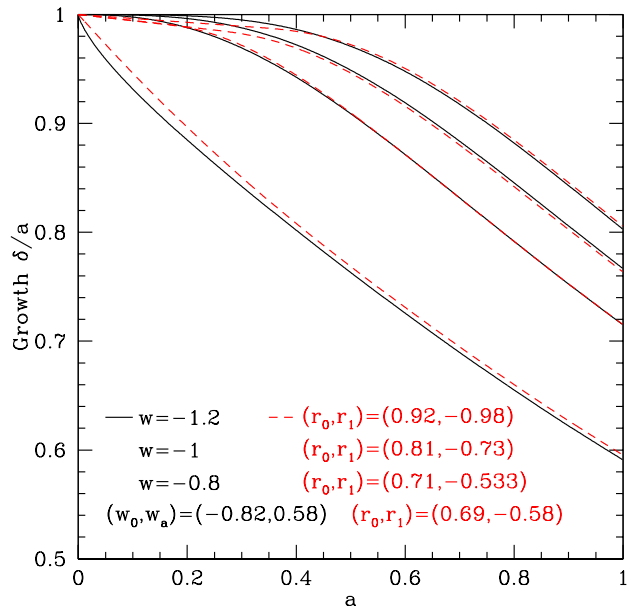,width=3.4in} 
\caption{Growth factor as in Fig.\ \ref{fig.grodhw}, but for the 
Ricci geometric dark energy models (red, dashed curves).  Simple 
parametrizations of these models can match 
the behavior of dark energy models (solid, black curves), including 
the cosmological constant. 
Slight deviations occur at higher redshifts. 
} 
\label{fig.grorhw}
\end{center} 
\end{figure}

Figure \ref{fig.grorhw} shows the growth curves for these four pairs of 
models.  At higher redshift the geometric models do have a deviation in 
growth behavior relative to the dark energy models, but this is small. 
Note that we enforce matter domination asymptotically, matching 
the $(r_0,r_1)$ parametrization onto $\calr=1/4$ at high redshift, but this 
is unlikely to be responsible for the deviation as the effect goes in the 
opposite direction, increasing the growth, and would enter at a different 
redshift than seen. 

\begin{figure}[!hbt]
\begin{center} 
\psfig{file=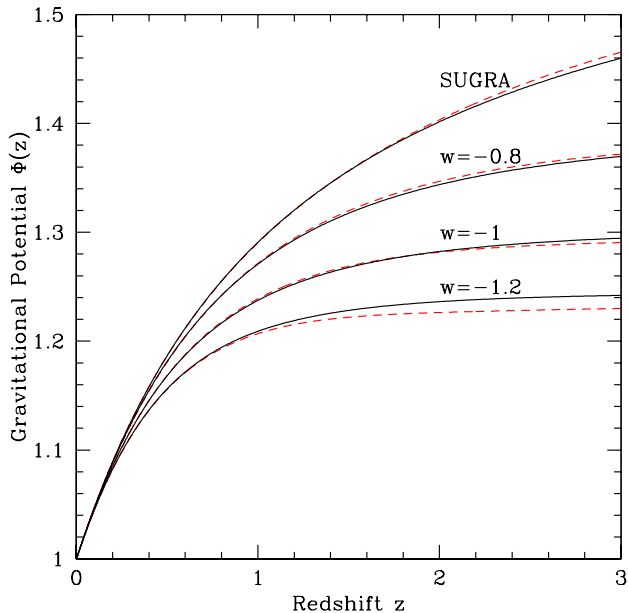,width=3.4in} 
\caption{The gravitational potential $\Phi(z)$ corresponding to the 
models of Fig. \ref{fig.grorhw}.  Slight deviations at higher redshifts 
occur between the Ricci models (dashed red curves) and their corresponding 
dark energy partners (solid black curves).  Deviations in {\it slope} 
focus on the behavior at $z\approx1-2$. 
} 
\label{fig.rhpotl}
\end{center} 
\end{figure}

The deviation can be seen more clearly in the gravitational potential 
decay behavior of Figure \ref{fig.rhpotl}.  Especially for the $w=-1.2$ 
case a distinction between the Ricci geometry and dark energy models 
can be seen, but this amounts to less than 1\% difference out to $z=3$. 
So for both cosmography and growth of structure, interpretation in terms 
of an effective equation of state remains a robust path, though not one 
that allows us to probe all the details of the fundamental physics 
responsible. 

Studying the behavior of the gravitational potential in Fig.\ 
\ref{fig.rhpotl} does offer one possible hope for elucidating the physics 
model in more detail.  At high redshift, $z\gg1$, we expect that all models 
approach the matter dominated behavior where the gravitational potential is 
constant.  This corresponds to the linear perturbation growth $\delta\sim 
a$.  Such behavior, of the development of structure through gravitational 
instability of adiabatic density perturbations, has been broadly successful 
in explaining the appearance of large scale structure in our universe.  In 
such a decelerating phase of the expansion, the origin of the accelerating 
physics should be largely moot. 

At low redshift, $z\ll1$, all the models within the region of parameter 
space our universe seems to inhabit show a similar behavior, all the curves 
of $\Phi(z)$ possessing nearly the same slope and so overlapping.  This does 
not arise from any fundamental requirement but is a coincidence for models 
with behavior not too different from a cosmological constant and for our 
universe at the present time, not too long after the acceleration began. 
(A similar coincidence makes the contours of constant age of the universe 
lie parallel to those of angular size corresponding to the first acoustic 
peak of the CMB, allowing for tight constraints on the age via CMB 
measurements \cite{knox}.) 

Since the slope of the gravitational potential-redshift relation is 
therefore fixed at the two ends (roughly 1/2 at $z=0$ and 0 at $z\gg1$), 
there will be some intermediate redshift where the deviation in slope 
$d\Phi/dz$ between models is maximal.  This in fact occurs when the 
dark energy or other accelerating mechanism begins to be dynamically 
significant, and the changing slope or curvature offers clues to the 
underlying physics, localized to this redshift. 

Certain cosmological observations relevant to the key redshift range of 
$z\approx1-2$ in fact are sensitive to this effect.  One is the integrated 
Sachs-Wolfe effect (ISW), where the CMB photon interaction with the time 
varying gravitational potential of large scale structure in the process 
of formation leads to CMB anisotropies on large angles or low multipoles. 
This involves $d\Phi/d\eta=H\,d\Phi/dz$ (see, for example, \cite{coohut}).  
Another prospective probe is the CMB bispectrum, related to the three point 
correlation function of temperature anisotropies arising from 
nongaussianities induced by weak gravitational lensing of the CMB by large 
scale structure (see \cite{spergelverde}).  This involves $\Phi\, 
(d\Phi/dz)$ and has been recognized to allow CMB measurements to have 
some sensitivity to the time variation of the dark energy equation of state 
\cite{giovibacci}, similarly localized to $z\approx1-2$.  Both these 
methods may be able to play a role in 
breaking the degeneracy between the physics of the spacetime geometry 
Ricci term and a physical dark energy.

\section{Superacceleration and the Big Rip} \label{sec.super} 

As alluded to in \S\ref{sec.accprin}, the value of the normalized 
Ricci scalar curvature $\calr=R/(12H^2)=1$ has a special role.  
The condition for superacceleration, where the acceleration 
increases with time, is $\calr>1$, which could be written $w_{\rm eff}<-1$. 
For the case of a physical dark energy component this implies that its 
energy density increases with expansion.  An important point regarding 
superacceleration is that it corresponds to $(\ddot a/a)\dot{}>0$ and not 
$(\ddot a)\dot{}>0$.  That is, the conformal acceleration is the relevant 
quantity.  

This is analogous to the condition for acceleration, or inflation, 
where $(aH)\dot{}>0$, meaning the conformal horizon $(aH)^{-1}$ shrinks with 
time.  Indeed such an acceleration condition is equivalent to $\dot H>-H^2$ 
while superacceleration relies on $\dot H>0$, equivalent to $(\ddot a/a) 
\dot{}>0$.  More explicitly, if $R<12H^2$ then $(\ddot a/a)<H^2$.  If this 
holds for all future times then $(\ddot a/a)\dot{}<(H^2)\dot{}=2H[(R/6)- 
2H^2]<0$.  Thus superacceleration is $(\ddot a/a)\dot{}>0$ and not 
$(\ddot a)\dot{}>0$.  The latter 
condition would be satisfied by a dark energy equation of state ratio 
$w<-2/3$, while $(\ddot a/a)\dot{}>0$ corresponds to $w<-1$.  

\begin{figure}[!hbt]
\begin{center} 
\psfig{file=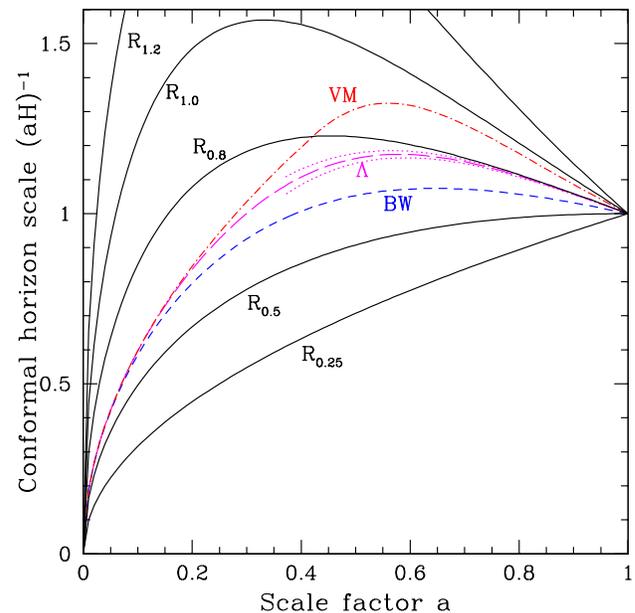,width=3.4in} 
\caption{The expansion history is plotted in terms of conformal horizon 
scale vs.\ scale factor for 
various modified gravity and spacetime geometry models.  The 
Ricci geometric dark energy models (solid, black curves) are subscripted 
with the present value $r_0$, and 
have the form $\calr=r_0+(1/4-r_0)(1-a)$.  All models are matter dominated 
in the past.  Negative slopes 
indicate an accelerating epoch while slopes more steeply negative than 
a critical value ($-1$ at the present) indicate superacceleration.  
} 
\label{fig.ahinv}
\end{center} 
\end{figure}

Figure \ref{fig.ahinv} illustrates the behavior of the conformal horizon 
in various cases, including those of Ricci geometric dark energy 
models listed by their present value of $\calr$.  Those shown follow 
eq.\ (\ref{eq.r01}) with constraint eq.\ (\ref{eq.rcon}). 
Any model with a region of negative slope is accelerating during 
such an epoch; e.g.\ the $r_0=0.5$ model is just starting to accelerate 
today, corresponding to $w_T=-1/3$.  The cosmological constant model has 
nearly the same acceleration today as for $r_0=0.8$, and $r_0=0.25$ 
is the (decelerating) Einstein-de Sitter cosmology.  Superacceleration 
requires a slope more steeply negative than $-(a^2H)^{-1}$, i.e.\ $-1$ today. 
This condition for superacceleration can be rewritten in terms of the 
logarithmic slope of the conformal diagram as 
\beq 
\frac{d\ln (aH)^{-1}}{d\ln a}<-1. 
\eeq 
It occurs for models steeper today than $r_0=1$, or more generally 
$\calr>1$ or $w_T<-1$.

From this diagram one can read off that a model such as vacuum 
metamorphosis is accelerating today but not superaccelerating.  
Although it acts as a component with $w<-1$, the total equation of 
state of the universe, including matter, is $w_T>-1$.  Even a 
currently superaccelerating model like $r_0=1$ only began accelerating 
at $z=2$, so we see that there is a relatively narrow range of 
redshifts -- not ``fine tuned'' -- when this extraordinary property 
of the universe will be evident.  

Note that such increasing conformal acceleration implies the existence of 
a Rindler horizon in the spacetime.  That is, points at a distance $r>1/g$ 
from an observer, where $g$ is the conformal acceleration, recede at 
greater than the speed of light and so are hidden behind a horizon 
\cite{rindler,davislineweaver}.   Generically such a horizon radiates 
particles at a temperature $T=g/(2\pi)$, analogous to Hawking radiation 
from a black hole horizon. 

Now we have seen that a component with $w<-1$, so-called phantom energy, 
leads to superacceleration.  This implies a Big Rip scenario for the 
fate of the universe, according to \cite{kamion}, where the increasing 
acceleration overcomes all other attractive forces.  However we conjecture 
that the particle creation from the Rindler horizon gives an energy density 
in radiation that grows faster than the phantom energy.  Illustratively, 
$\rho_R\sim T^4\sim (\ddot a/a)^4\sim \rho_{ph}^4$ while phantom energy 
dominates the universe.  So the ratio $\rho_R/\rho_{ph}\sim\rho_{ph}^3$ 
and this grows with time since $w<-1$.  Therefore at some point the 
radiation energy density will overtake the phantom energy density, 
shutting off the superacceleration.  Without superacceleration the 
particle creation declines, the radiation energy redshifts away, and the 
phantom energy can again dominate.  Depending on the details, this 
may lead either to an attractor at $w=-1$ or a cycle of superacceleration 
and hot, radiation (and matter) dominated phases of the universe.

\section{Conclusion} \label{sec.concl} 

To face the challenge of determining the fundamental physics 
responsible for the acceleration of the universe, we need to bring 
to bear next generation observations of the expansion 
history and possibly its dependent growth history.  The 
precision and 
accuracy of these future observations will guide us a long 
way to identifying new physics.  We see that at the heart of 
the next step lies a single function -- the effective equation 
of state $w(z)$.  Mapping this describes the cosmology; models with 
the same function, or equivalently same expansion history, will 
agree on the cosmological tests, whether distance-redshift, 
growth of structure, etc.  Furthermore the simple parametrization 
in terms of the present value, $w_0$, and a measure of the time 
variation, $w_a$, proves extraordinarily robust regardless of 
the exact reason for elaborating on the matter density term in the 
Friedmann equation. 

This is not to say there is no complementarity between 
cosmological probes; indeed that is a crucial ingredient in 
constraining the {\it values} of the equation of state 
parameters.  And next generation experiments will be superb 
at achieving this.  The simplicity of a two parameter functional 
form means we cannot easily appeal to ``naturalness'' 
to decide which physics model -- dark energy or modified 
gravity, say -- is a most likely explanation.  Despite the models 
considered here, though, there is no guarantee that an arbitrary 
modification $\delta H^2$ can be fit in terms of $w_0$, $w_a$. 
Regardless, the function $w(z)$ encodes all the standard, ``smooth'' 
information regardless of origin. 

We have illustrated this for several classes of physics 
including scalar field dark energy, modifications of general relativity in 
the Friedmann equation, and direct acceleration through 
Ricci ``geometric dark energy'', both in general and for 
specific models.  Explicit examples of the fits were given 
for probes such as magnitude-redshift, growth factor or 
gravitational potential, and distance to the CMB last 
scattering surface.  This held even for models with quite 
large time variation of the effective equation of state. 

One possible breakdown of the simple 
dark energy mimic ability might occur through the curvature 
of the gravitational potential decay behavior; the slope is 
remarkably model independent at low redshifts and asymptotically 
matter dominated at high redshift, but the localized deviation in 
between might provide a clue to the accelerating physics. 
Precision observations of the integrated Sachs-Wolfe effect 
or the lensing induced CMB bispectrum, yet untested, might 
be useful probes for this. 

We considered the implications of acceleration in general, 
regardless of origin, through the Ricci scalar curvature. 
This is pleasingly directly related to the expansion and fate 
of the universe.  In a conformal horizon history diagram 
(Fig.\ \ref{fig.ahinv}) we 
illustrate conditions for both acceleration and superacceleration, 
and briefly discuss the role of superacceleration in particle 
production that could nullify the Big Rip and indeed possibly 
provide an attractor for the universe to an apparent 
cosmological constant state. 

The picture of an achievable and wide ranging goal in measuring 
$w(z)$ is attractive. 
In our quest for understanding fundamental physics, though, we 
always want to push deeper.  The virtues of simplicity and 
broad applicability contest with lack of leverage in separating 
the root causes.  But it is only in the absence of new 
dynamics, new equations of motion, that the equation of state 
$w(z)$ or the expansion history $a(t)$ rules all.  New terms 
-- interactions or graininess -- lead to complexity but a grip on 
deeper details of the new physics.  This graininess could come from 
an observable consequence of dark energy perturbations or a 
noncanonical sound speed, separating it from a ``smooth'' gravity 
law (though it is only useful if it occurs within a realm 
accessible to precision observations).  Conversely, couplings 
in the gravitational sector, going 
beyond the Ricci spacetime geometry approach analyzed here, 
could distinguish a gravitational origin from one of dark 
energy.  This could arise in scalar-tensor theories, or 
metric perturbation terms $\dot h$ in the growth equation, 
or local curvature dependent effects $\delta R$, e.g. 
backreaction from structure formation. 

This is rather analogous to the situation in early universe 
acceleration -- inflation theory.  The incredible simplicity 
and generic power of it in solving cosmological and high energy 
physics conundra is 
immensely attractive, and we shouldn't lose sight of it, just 
as we shouldn't lose sight of the crucial role of $w(z)$.  But 
acceleration, then and now, is very much more than 
just a deSitter state.  We {\it want} complexity in the form 
of perturbations, tilt, gravitational waves to learn about 
the details of the fundamental physics.  For the CMB, 
measuring $\delta T/T$, or the 
power spectrum, is a stunning experimental accomplishment, 
just as $w(z)$ will be, but we want to explore further 
through nongaussianities, polarization, etc.  So too we look forward to 
probing gravity, dark energy, and acceleration.

\section*{Acknowledgments} 

This work has been supported 
in part by the Director, Office of Science, Department of Energy under 
grant DE-AC03-76SF00098.

\end{document}